\documentclass[aps,prb,twocolumn,superscriptaddress,preprintnumbers]{revtex4} %twocolumn
\usepackage[colorlinks,bookmarks=false,citecolor=blue,linkcolor=red,urlcolor=blue]{hyperref}
\usepackage{epsfig}
\usepackage{graphicx}
\usepackage{physics}

\usepackage{dcolumn}
\usepackage{bm}

\usepackage{amsmath,amssymb}

\makeatletter
\renewcommand*\env@matrix[1][*\c@MaxMatrixCols c]{%
  \hskip -\arraycolsep
  \let\@ifnextchar\new@ifnextchar
  \array{#1}}
\makeatother

\usepackage{appendix}

\newcommand{\bea}{\begin{eqnarray}}
\newcommand{\eea}{\end{eqnarray}}         

\begin{document}

\title{Disorder-induced currents as signatures of chiral superconductivity}

\author{Madhuparna Karmakar}
\email{madhuparna.k@gmail.com}
\affiliation{Department of Physics, Indian Institute of Technology Madras, Chennai 600036, India}
\author{R. Ganesh}
\email{ganesh@imsc.res.in}
\affiliation{The Institute of Mathematical Sciences, HBNI, C I T Campus, Chennai 600113, India}

\date{\today}

\begin{abstract}
Chiral superconductors are expected to carry a spontaneous, chiral and perpetual current along the sample edge. However, despite the availability of several candidate materials, such a current has not been observed in experiments. 
In this article, we suggest an alternative probe in the form of impurity-induced chiral currents. We first demonstrate that a single non-magnetic impurity induces an encircling chiral current. Its direction depends on the chirality of the order parameter and the sign of the impurity potential. Building on this observation, we consider the case of multiple impurities, e.g., realized as adatoms deposited on the surface of a candidate chiral superconductor. We contrast the response that is obtained in two cases: (a) when the impurities are all identical in sign
and (b) when the impurities have mixed positive and negative signs. 
The former leads to coherent currents within the sample, arising from the fusion of individual current loops. The latter produces loops of random chirality that lead to incoherent local currents. These two scenarios can be distinguished by measuring the induced magnetic field using recent probes such as diamond nitrogen-vacancy (NV) centres. 
We argue that impurity-induced currents may be easier to observe than edge currents, as they can be tuned by varying impurity strength, concentration and correlations. We demonstrate these results using a toy model for $p_x \pm i p_y$ superconductivity on a square lattice. We develop an improved scheme for Bogoliubov deGennes (BdG) simulations where both the order parameter as well as the magnetic field are determined self-consistently.  
\end{abstract}
% PACS, the Physics and Astronomy
                                 % Classification Scheme.
\maketitle
\section{Introduction}
A superconductor is said to be chiral if it spontaneously breaks time reversal symmetry\cite{Kallin2016_review,Wysokinski2019}. Such states have long been known, starting with liquid He$^3$ which is technically a superfluid and not a superconductor\cite{Leggett2006}. Interest in chiral superconductivity has grown alongside studies on non-superconducting time-reversal-symmetry-breaking states such as quantum Hall systems. In both classes of systems, typical excitation spectra show non-trivial topology. This shared thread has motivated extensive studies into edge currents. Despite several experimental studies and theoretical refinements, such a current has hitherto not been observed in a chiral superconductor. This situation can be contrasted with the relatively quick detection of edge states in topological insulators\cite{Nowack2013} and with the recent detection of an edge current in a topological non-chiral superconductor\cite{Wang2020}. In this article, we propose measuring a different type of current -- one induced by impurities. This provides a direct, observable and tuneable manifestation of chirality.

Several materials have been proposed as candidates for chiral superconductivity. The most prominent is Sr$_2$RuO$_4$, long thought to be a $p+ip$ superconductor. However, experimental studies have proved inconclusive. Recent studies have revived this debate\cite{Leggett2020}. Chiral superconductivity has also been proposed to occur in heavy fermion materials\cite{Maeno2012}, SrPtAs\cite{Biswas2013}, highly doped graphene\cite{Nandkishore2012,BlackSchaffer2014} and transition metal dichalcogenides\cite{Hsu2017,Ganesh2014}. In these materials, any evidence for chirality comes from bulk measurements such as $\mu$SR or the Kerr effect. In some cases, scanning tunnelling probes have found in-gap states at step edges\cite{Jiao2020}. However, an electrical current at an edge has not been observed so far\cite{Kirtley2007,Hicks2010}. 

In this backdrop, recent technologies to detect microscopic currents can break new ground. The most prominent of these uses nitrogen-vacancy (NV) centres -- atomic defects within a diamond probe. A spectroscopic measurement of excitations at the defect conveys information about the local magnetic field\cite{Casola2018}. As NV centres can operate over a wide temperature range and at nanometre separations from the sample, they can measure detailed field distributions. This has been used to map the spatial distribution of a current within graphene\cite{Tetienne2017,Ku2020}. With regard to edge currents of chiral superconductors, NV centres may succeed if they are sensitive enough to detect weak currents that were unobservable with earlier probes. However, edge currents may be unobservable due to deeper issues such as dissipation\cite{Marguerite2019}, disorder\cite{Ashby2009,Lederer2014} and compensating magnetisation textures\cite{Imai2012}. Such hurdles cannot be circumvented easily as edges are simple objects with no tuneable handles -- at least over coarse-grained length scales that are appropriate for mesoscopic probes. In this article, we describe an alternative signature where the magnitude and range of chiral currents can be tuned. We discuss signatures in magnetic field distributions that can be easily measured with probes such as NV centres. 

\section{Toy model for chiral superconductivity}

We consider a simple model with electrons hopping on a square lattice. The Hamiltonian is given by
\begin{eqnarray}
H = -t \sum_{\sigma,\langle ij \rangle} \left\{ e^{i \frac{e}{\hbar} \mathcal{A}_{ij}}c_{i,\sigma}^\dagger c_{j,\sigma} + h.c. \right\} + H_{int},
\label{eq.model}
\end{eqnarray} 
We restrict hoppings to nearest neighbours for simplicity. The hopping amplitude, $t$, sets the energy scale for the problem. Each hopping element carriers a Peierls' phase that encodes the vector potential, with $\mathcal{A}_{ij} = \int_i^j \vec{A}(\mathbf{r})\cdot \mathbf{dl}$. We have denoted the charge of the electron as $e$. 
The vector potential here originates from supercurrents in the sample and not from an externally imposed magnetic field. Its precise form will be discussed below. 

We choose an artificial interaction term designed to give rise to chiral superconductivity,
\begin{eqnarray}
H_{int} = -g \sum_{\langle ij \rangle} \mathcal{D}_{ij}^\dagger \mathcal{D}_{ij}.
\label{eq.Hint}
\end{eqnarray}
The sum runs over all nearest neighbour pairs of sites. The coupling constant, $g$, is taken to be positive. We introduce a two-site Cooper pair annihilation operator, 
\begin{eqnarray}
\mathcal{D}_{ij} = c_{i,\mu}(\sigma_3 i \sigma_2)_{\mu,\mu'} c_{j,\mu'} =  \{ c_{i,\uparrow} c_{j,\downarrow}  + c_{i,\downarrow} c_{j,\uparrow} \},
\end{eqnarray}
where the $\sigma$'s are Pauli matrices. 
While a generic triplet Cooper pair can contain three components with spin $m_z = -1,0,1$, we have only retained the $m_z = 0$ component in $\mathcal{D}_{ij}$. In order to gain energy from the interaction term, the Cooper pair operator acquires a non-zero expectation value with $\langle \mathcal{D}_{ij} \rangle = \Delta_{ij}$, a complex scalar. However, this expectation value must satisfy certain constraints. From fermionic anticommutation relations, we see that $\Delta_{ij} = -{\Delta}_{ji}$. This signifies that the order parameter has $p$-wave symmetry, changing sign under a $\pi$-rotation. For instance, upon a $\pi$ rotation about a bond centre, the bond direction changes from $ij$ to $ji$ so that the order parameter flips sign. Each site has four bonds emanating from it. In order to lower energy on each of these bonds, the order parameter naturally develops chiral $p_{x}\pm i p_{y}$ symmetry. Moving counterclockwise around a site, the phase of the order parameter changes by $\pm \pi/2$ from one bond to the next. 

A detailed study of this model can be found in Ref.~\onlinecite{Clara2017}. In summary, the system spontaneously chooses between one of two chiralities, i.e., between $p_x + i p_y$ and $p_x - i p_y$ character. Ordering introduces a quasiparticle gap in an infinite system (or in one with periodic boundaries), with the same spectrum for either choice of chirality. However, in the presence of an edge, we find a sharp signature that distinguishes the two chiralities. The quasiparticle spectrum hosts a mode within the bulk gap that propagates along the edge. The direction of propagation is determined by the chirality of the bulk order parameter. It is this mode that carries the edge current, providing a signature of chiral superconductivity. As a quasiparticle mode in a superconductor, this mode is a mixture of particle and hole components. As a consequence, it does not carry a quantized charge current -- unlike an edge current in a Chern insulator. 

\subsection{Role of impurities}
We now introduce impurities into the problem in the form of potentials on a certain fraction of sites. For simplicity, 
we assume that each impurity generates a local potential with a fixed amplitude, denoted as $W$. 
As for the sign of the impurity potential, we consider two schemes: (i) correlated disorder -- with every impurity carrying the same sign, and (ii) random disorder -- with each impurity having a random sign. We represent these as 
\begin{eqnarray}
H_{corr.} &=& W\sum_{\{i\}} \sum_\sigma c_{i,\sigma}^\dagger c_{i,\sigma}, 
\label{eq.Hcorr} \\
H_{rand.} &=& W\sum_{\{i\}} \delta_i\sum_\sigma c_{i,\sigma}^\dagger c_{i,\sigma}.
\label{eq.Hrandom}
\end{eqnarray}
In both cases, the sum over $i$ runs over a randomly selected subset of sites. We consider the fraction of impurity sites to be a tuneable parameter, $\eta$. In the random case, $\delta_i = \pm 1$ is an Ising-like random variable that is independently chosen at each impurity site. These two impurity schemes can be realized by placing a collection of adatoms on the surface of a chiral superconductor. If all adatoms are of the same type, this corresponds to the situation encoded in $H_{corr.}$. Here, the sign of the potential is decided by the relative electronegativity of the adatom with respect to the substrate. On the same lines, deposition of two species of adatoms can realize $H_{rand.}$.
 
\section{Bogoliubov deGennes simulations}

We consider the system on a $L\times L$ lattice with periodic boundary conditions. We perform a mean-field decomposition of the interaction term in Eq.~\ref{eq.Hint}, by defining bond variables $\Delta_{ij} \equiv \langle \mathcal{D}_{ij} \rangle$. This allows us to write the Hamiltonian as a quadratic form with 
\begin{eqnarray}
\hat{H} = \Psi^\dagger M \Psi,
\label{eq.Hquad}
\end{eqnarray}
where $\Psi = \{ \ldots c_{i,\uparrow}\ldots c_{i,\downarrow}^\dagger \ldots \}^T$. The $2L \times 2L$ Hamiltonian matrix can be written as 
\begin{eqnarray}
M = \left( \begin{array}{cc}
K(t,\vec{A},W) & P( \Delta_{mn} ) \\
P( \Delta_{mn} )^\dagger & -K(t,\vec{A},W)^T
\end{array} \right).
\end{eqnarray}
The diagonal blocks, $K(t,\vec{A},W)$ encode hoppings with Peierls' phases and impurity potentials. The off-diagonal blocks contain the mean-fields, with $\Delta_{mn}$ defined on the nearest-neighbour bond connecting sites $m$ and $n$. In typical Bogoliubov deGennes (BdG) calculations, the superconducting order parameters are determined using $\Delta_{mn} \equiv \langle \mathcal{D}_{mn} \rangle$. At the same time, the vector potential is either ignored or assumed to be fixed by a strong (non-self-consistent) external field. In this article, we present an improved BdG scheme where the vector potential is also determined self-consistently. 

To determine the vector potential, we note that each bond carries a current. By adapting the arguments in 
Ref.~\onlinecite{Scalapino1993}, we define the bond current operator as ${J}_{mn} = \langle (\partial{\hat{H}}/\partial {\mathcal{A}_{mn}} )\rangle$, where $\hat{H}$ is the Hamiltonian of Eq.~\ref{eq.Hquad} and $<\cdot >$ represents the expectation value. Here, $(m,n)$ represents a pair of neighbouring sites. We obtain
\begin{eqnarray}
 {J}_{mn} &\approx &  \frac{e}{\hbar}\langle \hat{\mathcal{J}}_{mn} \rangle - \left(\frac{e}{\hbar}\right)^2 \langle \hat{\mathcal{K}}_{mn} \rangle \mathcal{A}_{mn} +\mathcal{O}(\mathcal{A}_{mn}^2), 
\label{eq.currentexp}
\end{eqnarray}
where
\begin{eqnarray}
\hat{\mathcal{J}}_{mn} &=&  -t \sum_{\sigma}\{ i c_{m,\sigma}^\dagger c_{n,\sigma} - i c_{n,\sigma}^\dagger c_{m,\sigma}
\}, \\
\hat{\mathcal{K}}_{mn} &=& -t \sum_{\sigma} \{ c_{m,\sigma}^\dagger c_{n,\sigma} +  c_{n,\sigma}^\dagger c_{m,\sigma}\} .
\label{eq.bondcurrent}
\end{eqnarray}
In the absence of a vector potential, $\hat{\mathcal{J}}_{mn}$ and $\hat{\mathcal{K}}_{mn}$ would encode the bond current and the bond energy respectively. We neglect $\mathcal{O}(\mathcal{A}_{mn}^2)$ terms here, assuming that $\mathcal{A}_{mn}$ is small on every bond. This is justified in the limit of weak disorder. Currents in our system are induced solely by disorder. Weak disorder will give rise to weak currents and, in turn, to a weak vector potential. 

We next discuss the vector potential induced by the bond currents of Eq.~\ref{eq.bondcurrent}. Assuming the Coulomb gauge condition ($\vec{\nabla} \cdot \vec{A} =0 $), a closed form expression can be derived for the vector potential induced by a set of static source currents in two dimensions,
\begin{eqnarray}
\vec{A}(\mathbf{r}) = \frac{\mu_0}{4\pi} \int d^2 r' \frac{\vec{J}(r')}{\vert \mathbf{r} - \mathbf{r}' \vert},
\label{eq.Avec}
\end{eqnarray} 
where $\vec{J}(r')$ is the current density, i.e., current per unit cross-sectional length. 
This relation follows from the Biot-Savart law and the integral solution to the Poisson's equation.  
We adapt this equation to our lattice problem using
\begin{eqnarray}
\mathcal{A}_{ij}= \frac{\mu_0}{4\pi} \ell^2 \left[ \sum_{\{kl\} \parallel \{ij\}} \frac{J_{kl}}{\vert \mathbf{R}_{kl} - \mathbf{R}_{ij} \vert} + \frac{J_{ij}}{\delta \ell}\right].
\label{eq.Aveclattice}
\end{eqnarray} 

Here, $\ell$ represents the lattice constant. The index ${ij}$ represents the reference bond where the vector potential is to be determined. The position of the centre of the bond is denoted as $\mathbf{R}_{ij}$. The sum runs over ${kl}$, all \textit{other} bonds that are parallel to ${ij}$, i.e., if ${ij}$ is a bond in the $x$-direction, $kl$ runs over all $x$-bonds except ${ij}$. The centre of each of these bonds is denoted as $\mathbf{R}_{kl}$. We calculate inter-bond distances ($\vert \mathbf{R}_{kl} - \mathbf{R}_{ij} \vert$), in units of $\ell$, the lattice constant. The last term in the above equation represents the contribution from the very same bond, $ij$. Naively this represents a singular contribution as the source current and the target bond coincide, i.e., $\vert \mathbf{R}_{kl} -\mathbf{R}_{ij} \vert$ vanishes. 
However, this contribution will be cut off by microscopic length scales which can be visualized as follows. We may view each bond as being composed of a bundle of wires, with each wire corresponding to an orbital overlap. If the wires are slightly displaced from the bond vector, the singular contribution is regularized. We phenomenologically model such effects using a cutoff length scale denoted as $\delta \ell$.  

We obtain the order parameters and the vector potential self-consistently. Starting with random values of $\Delta_{ij}$'s and $\mathcal{A}_{ij}$'s, we obtain a new set of values as described above. We iterate this process until self-consistency is achieved. The calculations are carried out while respecting periodic boundary conditions. Each distance ($\mathbf{R}_{ij} - \mathbf{R}_{kl}$) in Eq.~\ref{eq.Aveclattice} is chosen in accordance with the periodicity of the lattice. For example, two bonds are placed in neighbouring unit cells if that gives the lowest distance between them. We note that our approach does not give consistent results with open boundary conditions. In the presence of an edge, the vector potential outside the sample cannot be taken into account. This prevents us from ensuring $\vec{\nabla}\cdot \vec{A}=0$, a necessary condition for the use of Eq.~\ref{eq.Avec}. 
 
We emphasize that our scheme is an improvement over standard BdG approaches. As the vector potential is determined self-consistently, our approach allows for `screening' currents. These can be viewed as a manifestation of the Meissner effect, with any magnetic field diminished by counter-currents that naturally arise in a superconductor. In the context of Sr$_2$RuO$_4$, it has been shown that edge currents are indeed strongly screened\cite{Matsumoto1999,Furusaki2001}.

\subsection{Parameters in the simulation}

In the BdG scheme outlined above, the input parameters are the hopping strength ($t$) and the coupling constant ($g$). The vector potentials on bonds, $\mathcal{A}_{ij}$'s, are the desired output as they immediately yield the magnetic flux through each plaquette. 

The currents and the vector potentials require particular attention as they are accompanied by dimension-carrying constants. They can be determined in a consistent fashion as follows. We first assert that the bond currents in Eq.~\ref{eq.bondcurrent} are proportional to $( te/\hbar)$, a combination that carries dimensions of current. Operationally, we determine the currents up to this factor. Similarly, we calculate the vector potentials in Eq.~\ref{eq.Aveclattice} up to $(\mu_0 \ell / 4\pi) (te/\hbar)$, a combination that carries dimensions of the line integral of the vector potential ($\int \vec{A}\cdot \vec{dl}$). We have a lower power of $\ell$ as compared to the prefactor in Eq.~\ref{eq.Aveclattice} as the lattice distances ($\vert \mathbf{R}_{kl} - \mathbf{R}_{ij} \vert$) are also proportional to $\ell$. The factor of $(te/\hbar)$ comes from the current term ($J_{kl}$) on the right hand side of Eq.~\ref{eq.Aveclattice}. 

To determine currents and vector potentials, we require numerical values for two dimensionless parameters:
\begin{enumerate}
\item $\zeta \equiv (\mu_0 \ell / 4\pi) (te^2/\hbar^2)$: this quantity appears in two places. The first is in the Peierls' phases of Eq.~\ref{eq.model} where the vector potential, $\mathcal{A}_{ij}$, is multiplied by $e/\hbar$. The second is the current expression of Eq.~\ref{eq.bondcurrent}, as a prefactor to the second term on the right hand side. This quantity, $\zeta$, contains two material-dependent parameters $t$ and $\ell$, apart from fundamental constants. To estimate its numerical value,  
we take $\ell = 10\AA$ and $t\sim 2.5 eV$ so that $\zeta \sim 10^{-4} $.  
\item $\delta \ell/\ell$: this quantity encodes the regularization parameter in Eq.~\ref{eq.Aveclattice}. In our calculations, we assume an ad hoc value for this parameter, $\delta \ell / \ell = 0.1$. 
\end{enumerate}

\section{Response to a single impurity}

We first consider the case of a single impurity in the chiral superconductor. Before discussing our results, we review some basic ideas. Chirality allows for a spontaneous local current that circulates around an impurity. This can be understood as follows. We imagine a hole cut into the superconductor. As the superconductor is chiral, it hosts a chiral boundary current that propagates along the edge. The hole can be replaced by a plateau-shaped potential which repels electrons from the interior region. We now imagine shrinking the radius of the `hole', with the boundary current smoothly deforming to cling to the shrinking perimeter. When the radius approaches zero, we are left with a point-like potential. In this limit, the boundary current takes the form of a local current encircling the impurity. The current is no longer quantized as we can now have scattering between opposite edges of the hole. Nevertheless, we expect a non-zero current strength. Based on similar arguments, a impurity-induced current has been studied in a Chern insulator\cite{Jha2017}.

Before presenting results on impurity response, we discuss chirality in our mean-field solutions. In clean as well as in disordered systems, we find two mean-field solutions that correspond to a choice of chirality. They can be understood as $\Delta_x \pm i \Delta_y$ configurations: the order parameters on the $x$ and $y$ bonds differ by a phase of $\pm \pi/2$. The chirality of the configuration is determined by the choice of initial values in our BdG scheme. If the initial values are chosen to be unbiased towards either chirality, the system picks one of the two chiralities at random. For the situations considered in this article, we did not find `domain-wall' solutions containing regions of differing chirality. For concreteness, in the discussion below, we restrict our attention to solutions of the form $\Delta_x + i \Delta_y$.  

\begin{figure*}
\includegraphics[width=6in]{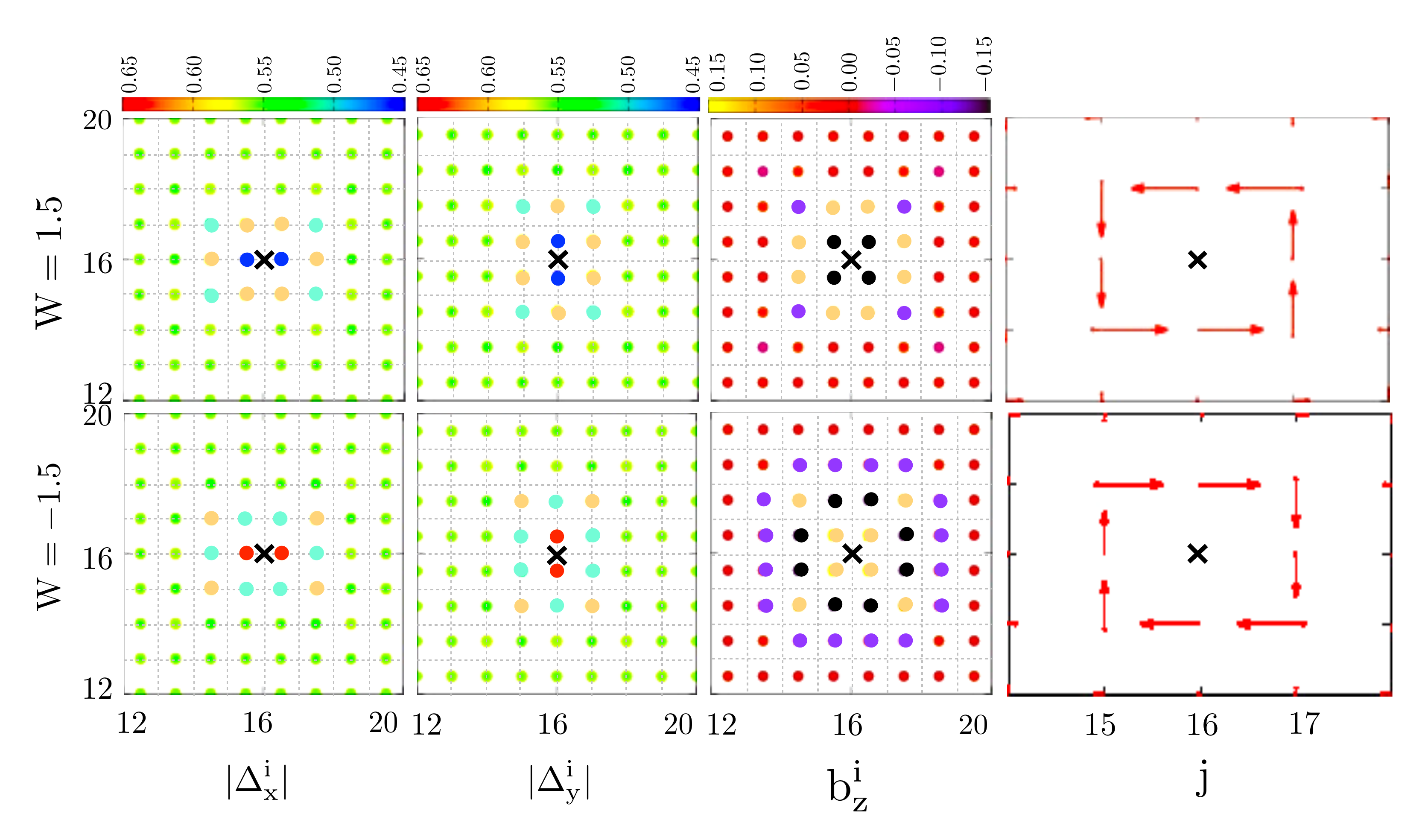}%{single_imp_maps_new2}
\caption{(Colour online) Response to a single impurity corresponding to a positive (top panels) and a negative (bottom panels) on-site potential. The data have been calculated for $g=4t$ with the impurity position indicated by an `x'. The chirality in all panels is of the $\Delta_x + i \Delta_y$ type. From left to right, the panels show the order parameter on the x-bonds, those on the y-bonds, the magnetic field through each plaquette and the induced bond currents.  }
\label{fig.single_imp}
\end{figure*}

Our results for the response to a single impurity are shown in Fig.~\ref{fig.single_imp}. The panels on the left show the order parameter amplitude on $x$ bonds. The panels in the second column show amplitudes on the $y$ bonds. Here, $\Delta_{x/y}^i$ is the order parameter on the bond emanating from site $i$ in the positive $x/y$-direction. 
In these plots, we see that the impurity creates a local distortion in order parameters. 
In addition, magnetic fields are generated as shown in the third column. The fields (fluxes, to be more precise) are defined on plaquettes, given by the sum of Peierls' phases (vector potentials) along four bonds. The field is strongest at the plaquettes that are immediately adjacent to the impurity, weakening as we move further. Crucially, the direction of magnetic flux on a given plaquette is determined by the sign of the impurity potential.
This field originates from a current that circulates around the impurity, as shown in the panels on the right. The current is only appreciable on bonds that are close to the impurity. For a given background chirality, the direction of the current loop is set by the sign of the impurity potential. 

\begin{figure}
\includegraphics[width=\columnwidth]{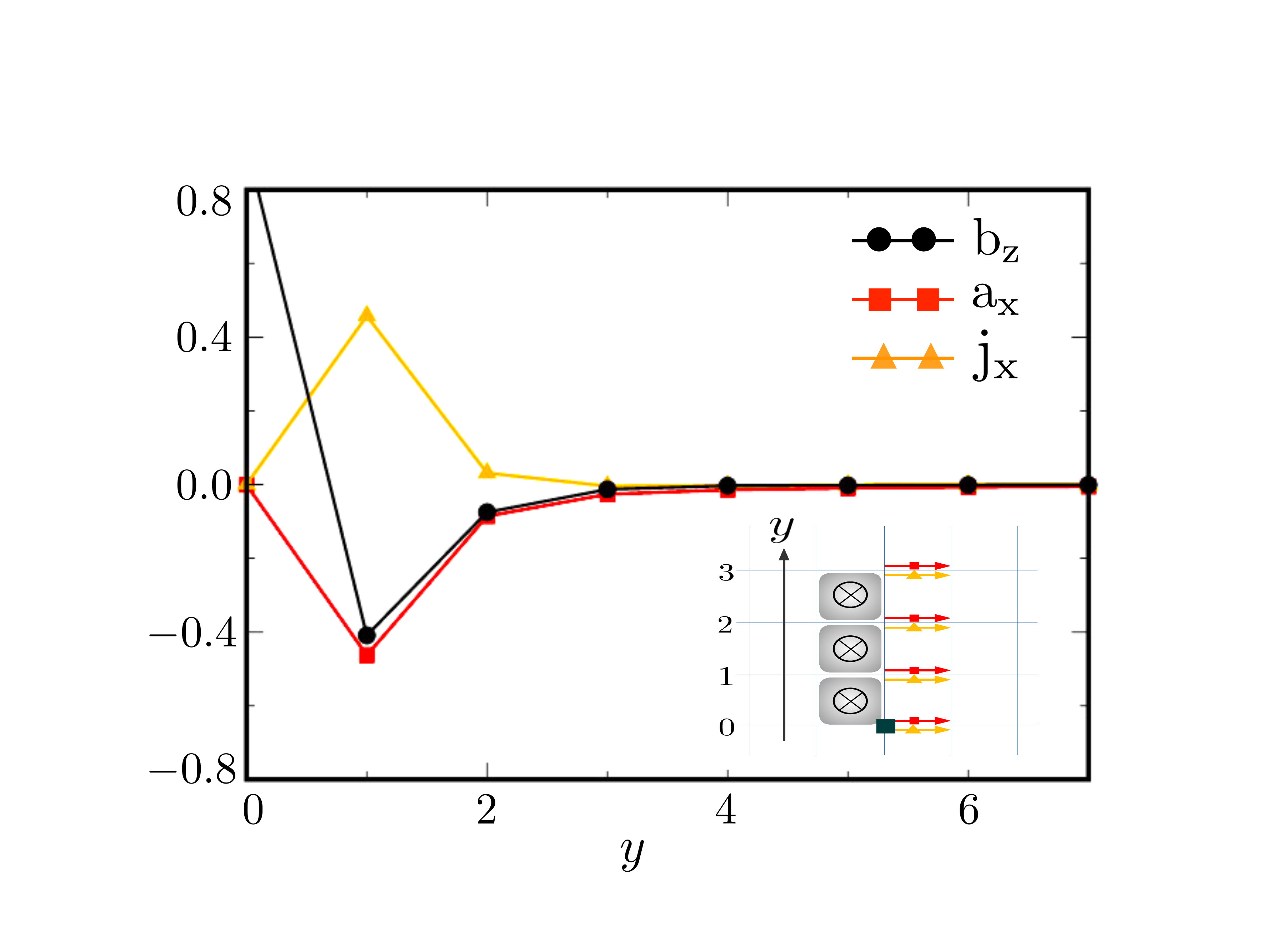} %{single_imp_profile_W3p5.pdf}
\caption{(Colour online) Response to a single impurity. Profiles of induced current, vector potential and magnetic field for $W=3.5t$ and $g=4t$. In the enclosed cartoon, the impurity location is shown as a dark square. This point is taken to be $(0,0)$. The current and vector potential are shown along bonds parallel to the $x$-axis at various values of $y$. The field shown is on plaquettes at various $y$ values.}
\label{fig.imp_profile}
\end{figure}

In Fig.~\ref{fig.imp_profile}, we show the profile generated by a single impurity. We show the current induced on neighbouring bonds, the resulting vector potential and the magnetic field. We see that the response is highly localized, decaying sharply as we move away from the impurity. This figure is to be compared with well-known results on edge currents, e.g., in Ref.~\onlinecite{Furusaki2001}. Our problem is fundamentally different as it lacks the translational symmetry of a linear edge. Nevertheless, the profiles show some degree of similarity. This comparison also brings out an advantage of impurities over edges. An edge is a sharply defined feature with no additional tuning parameters (apart from deforming the edge shape).  In contrast, an impurity is inherently tuneable as both the sign as well as the amplitude of the impurity potential can be varied. This allows for tunability of the response. For example, the direction of the current can be switched by simply changing the sign of the potential. 

\section{Response to multiple impurities}
We next consider the response to multiple impurities. 
\subsection{Correlated disorder}
\begin{figure}
\includegraphics[width=\columnwidth]{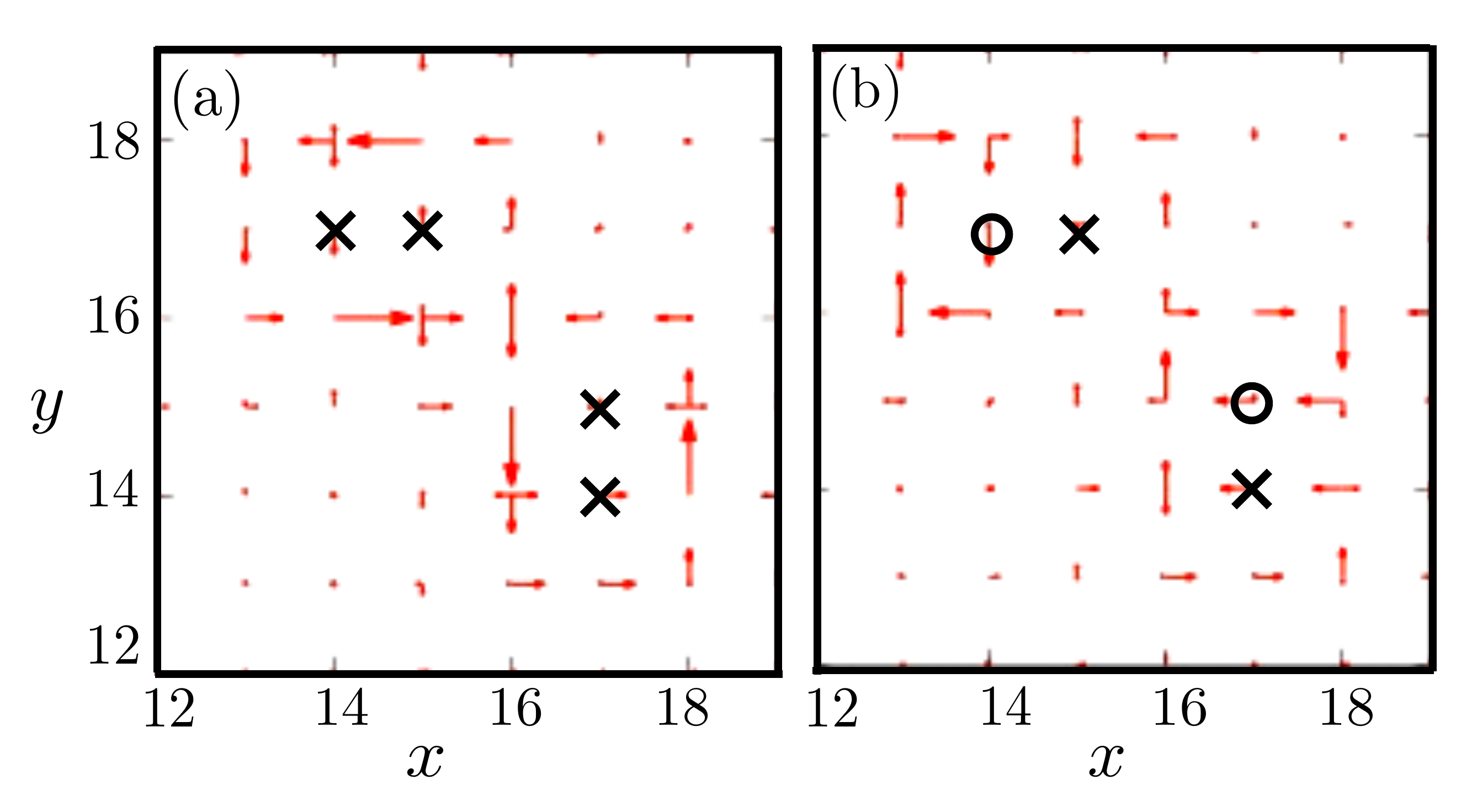} %{current_loops}
\caption{(Colour online) Currents with proximate impurities for $W=0.5t$ and $g=4t$. In (a), all impurities have 
the same sign. Their positions are labelled with `x'. In (b), impurities have mixed sign, indicated using `x' and `o'.}
\label{fig.loop_corr}
\end{figure}

We first take up the case of correlated impurities, as defined in Eq.~\ref{eq.Hcorr}. As discussed in the previous section, each impurity induces a local current loop whose chirality is determined by the sign of the impurity potential. As all impurities are of the same sign, we have many current loops with the same chirality. As a result, where impurities are close to one another, their currents fuse to form a larger loop as depicted in Fig.~\ref{fig.loop_corr}(left). 

\begin{figure*}
\includegraphics[width=6in]{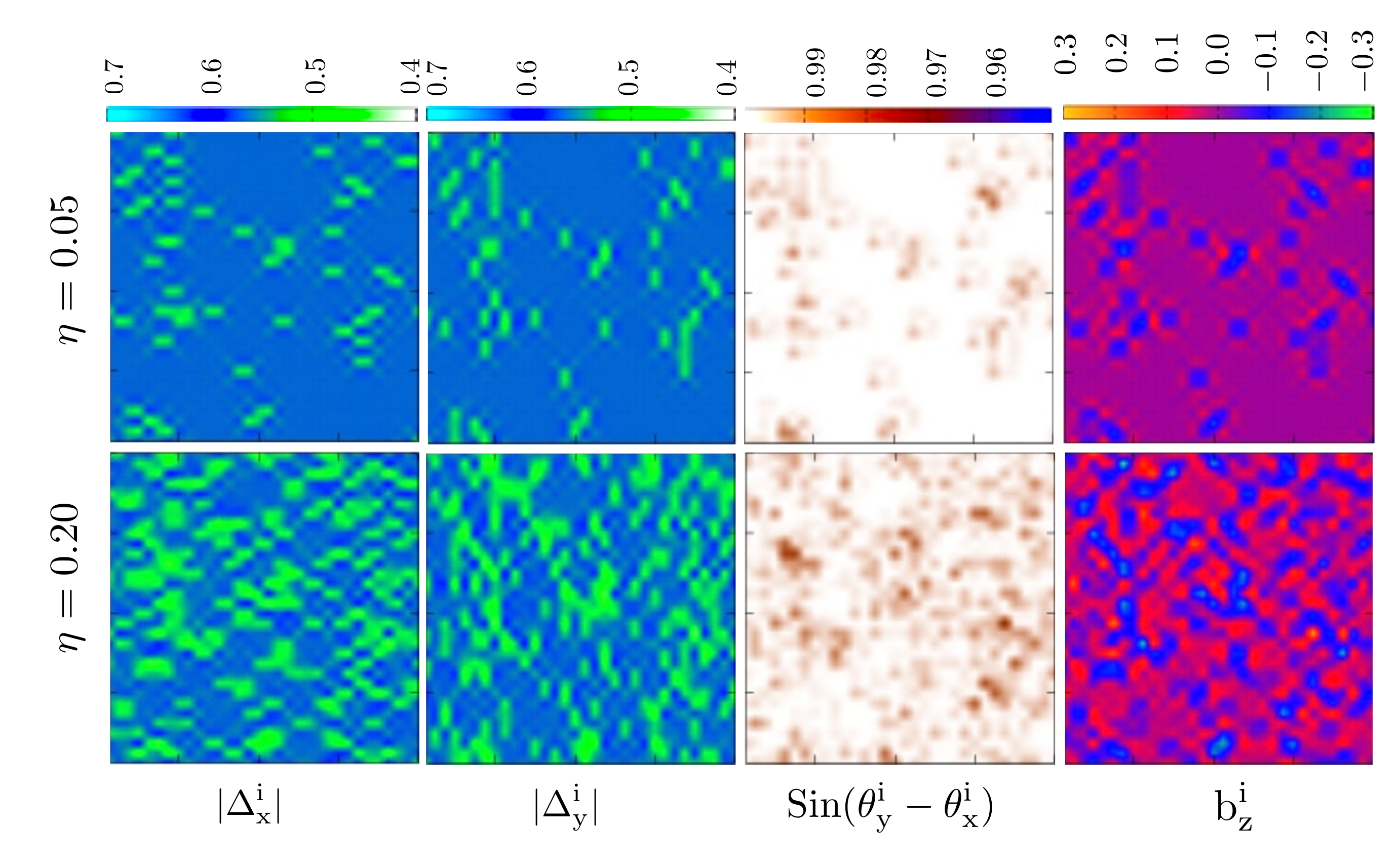} %{multi_imp_correlated_new_maps}
\caption{(Colour online) Response to correlated impurities with $W=0.5t$ and $g=4t$ on a $32 \times 32$ lattice. We show snapshots for two disorder realizations with concentrations $\eta = 0.05$ and $\eta = 0.2$. From left to right, we show order parameter amplitudes on $x$ bonds, those on $y$ bonds, sine of the phase difference in the two components and finally, the magnetic field. }
\label{fig.impcorr}
\end{figure*}

The fusing of current loops lead to observable macroscopic consequences. In Fig.~\ref{fig.impcorr}, we show physical properties for two disorder realizations with concentrations $\eta = 0.05$ and $\eta=0.2$. As the order parameters are complex scalars, we write them as $\Delta_{x/y}^j = \vert \Delta_{x/y}^j \vert e^{ i \theta_{x/y}^j}$, where $j$ is a site index and the $\theta_{x/y}^j$ is the order parameter phase. In the panels on the left, we show the amplitudes of the order parameters, $\vert \Delta_x \vert $ and $\vert\Delta_y\vert$. We see that impurities locally suppress the amplitudes of both $\Delta_x$ and $\Delta_y$ components. This is a consequence of the positive sign of the impurity potential. When the negative sign is used, we see local enhancement of order parameters. In the third panel, we plot $\sin(\theta_x - \theta_y)$. Here, $\theta_x$ is the phase of $\Delta_x$ while $\theta_y$ is that of $\Delta_y$. The sine of the difference in angles is a measure of chirality. In the $\Delta_x \pm i \Delta_y$ states, we expect $(\theta_x - \theta_y)=\pm \pi/2$. The sine of this quantity takes the value of $\pm 1$, serving as an indicator of chirality. In the figure, we see that chirality is robust to impurities. Within the parameter regime that we have explored, we do not see any domain walls separating regions of differing chirality.  

In the fourth column in the figure, we show the spatial distribution of the magnetic field. At low impurity concentration, we see localized (negative) magnetic fields around each impurity. At larger impurity concentrations, these field regions come closer to one another. In the figure, for the two concentrations shown, we see field distributions that are skewed towards negative values. This is tied to the sign of the impurity potential. The spread is set by $W$, the magnitude of each impurity potential, as well as the impurity concentration, $\eta$. This field distribution is a measurable signature of impurities that can be accessed by probes such as NV centres.

\subsection{Random disorder}
\begin{figure*}
\includegraphics[width=6in]{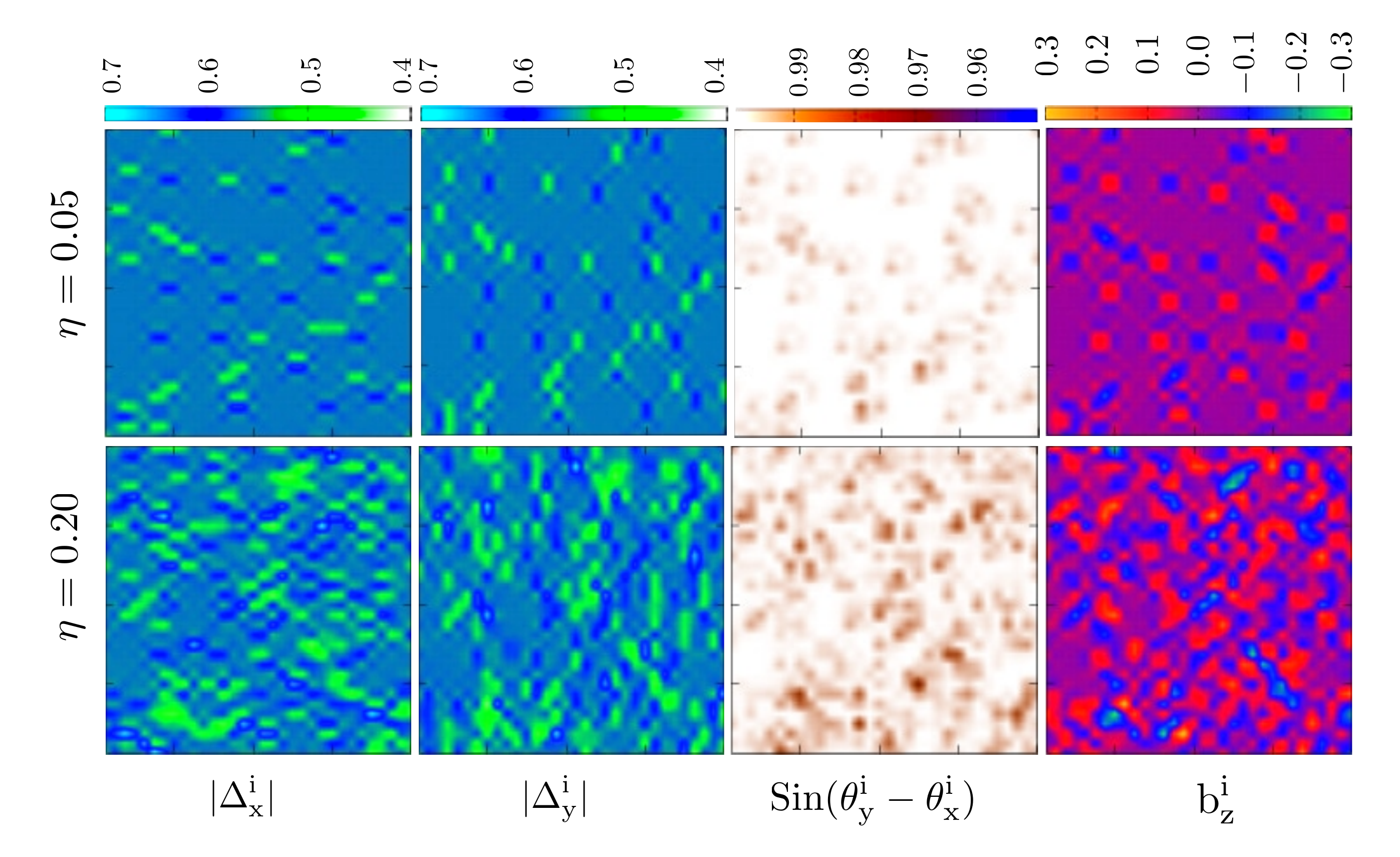} %{multi_imp_random_new_maps}
\caption{(Colour online) Response to random impurities. The parameters and the plotted quantities are the same as in Fig.~\ref{fig.impcorr}. }
\label{fig.imprand}
\end{figure*}

We next consider disorder with random sign as given by Eq.~\ref{eq.Hrandom}. Here, impurity-generated current loops have random chiralities. As shown in  Fig.~\ref{fig.loop_corr}(right), this leads to incoherent currents with no loop structure. We obtain strong local currents in random directions. In Fig.~\ref{fig.imprand}, we show physical properties that arise from random disorder. In the left two panels, we show the amplitudes of the order parameters. These plots are to be compared with the corresponding panels in Fig.~\ref{fig.impcorr} where order parameters were suppressed at impurity sites. Here, we see suppression in some regions and enhancement at others. The suppression or enhancement directly correlates with the sign of the impurity potential. As the impurity potential changes sign from one site to another, the response in the amplitude changes as well.

In the third panel, we plot $\sin(\theta_x - \theta_y)$, an indicator of chirality as described above. The chirality remains robust with no sign of domain formation. This is similar to the case of correlated disorder, shown in Fig.~\ref{fig.impcorr}. We note that this robustness of chirality to disorder may only hold in our specific model and regime of interest. In other systems, it is conceivable that disorder (especially in the form of magnetic impurities) may lead to domain formation. 

The fourth column depicts the induced magnetic flux. Each impurity gives rise to a local flux, with the flux direction fixed by the sign of the impurity. Away from impurities, the flux is close to zero. Around a positive or negative impurity, the flux takes a negative or positive value, respectively. This leads to a broad distribution with roughly equal spread on either side of zero. The width of the distribution is set by $W$, the strength of the impurity potential, as well as $\eta$, the impurity concentration. The field distribution here is visibly different from that induced by correlated disorder, especially at low impurity concentrations. In the correlated case, as seen in Fig.~\ref{fig.impcorr}, impurities generate local flux peaks that have the same sign. With random disorder, we see peaks of both signs.
%, correlated with the sign of the impurity potential. 

\begin{figure}
\includegraphics[width=\columnwidth]{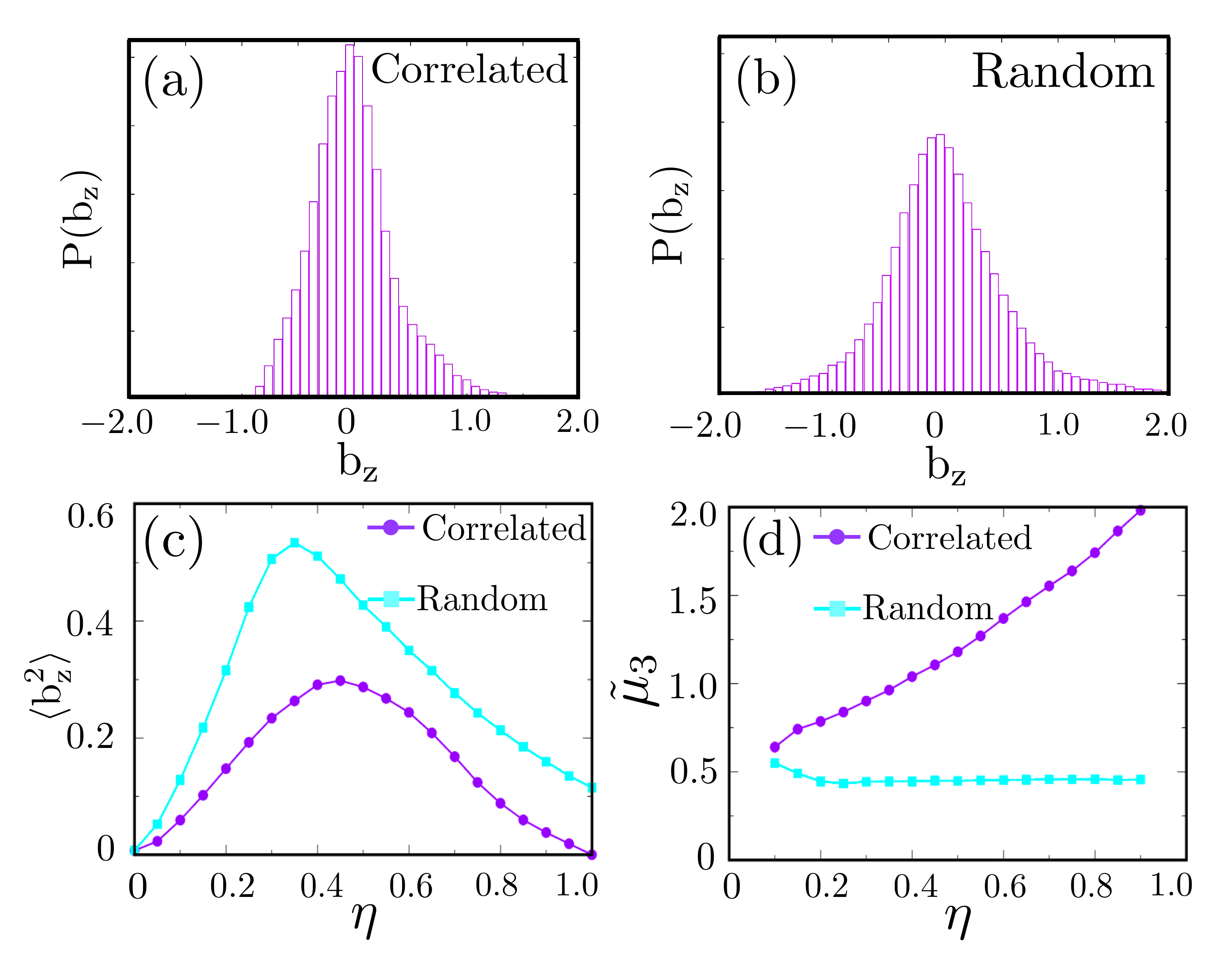} %{hist_skew_combo_W2}
\caption{(Colour online) Magnetic field distributions generated by correlated and random disorder. Histograms of field values are shown in (a) and (b), with the disorder concentration fixed at 20\%. (c) The variance of the distributions is shown as a function of disorder concentration. (d) The skewness of the distributions vs. disorder concentration. In all panels, the data corresponds to $g=4t$, $W=2t$ and $L=32$, averaged over 100 disorder realizations.}
\label{fig.Bsq}
\end{figure}

\subsection{Comparing field distributions}
We focus on magnetic field distributions, as they are particularly suited for measurements by $NV$ centre probes. In Fig.~\ref{fig.Bsq}, we compare the field distributions generated by correlated and random disorder. In panels (a) and (b), we show histograms of field values. In both cases, the mean of the distribution is pinned at zero, i.e., $\langle b_z \rangle =0$. This is a consequence of periodic boundary conditions, whereby our lattice system forms a torus. As this torus does not enclose a magnetic monopole charge, the net magnetic flux must vanish and the mean of the field distribution must remain zero. Despite having the same mean value, the two field distributions are markedly different. 

Correlated disorder creates a skewed distribution as each impurity induces a current loop with the same chirality. The distribution has a longer tail on one side as seen from Fig.~\ref{fig.Bsq}(a). This reflects a bias towards one magnetic field direction over the other. In contrast, random disorder leads to local currents with both chiralities. This leads to a nearly symmetric field distribution centred at zero. The distribution is wider than in the correlated case. This is a consequence of having strong local currents when impurities with opposite signs are in close proximity. These lead to localized regions with strong fields. Such regions can correspond to either of the two field directions. 

We compare quantitative measures of these distributions in panels (c) and (d). Random disorder leads to a higher variance for a given disorder concentration. This reflects a wider spread in field values, as can be seen visually by comparing the histograms in panels (b) and (a). In panel (c), we see that the variance in the field distribution changes non-monotonically with disorder concentration. This is because strong disorder suppresses order parameter amplitudes and, in turn, the supercurrents in the sample. 

We next compare the skewness of the distributions, defined as $\tilde{\mu}_3 = E[\{(X-\mu)/\sigma\}^3]$. This quantifies the asymmetry of the distribution about the mean. As seen in Fig.~\ref{fig.Bsq}(d), correlated disorder leads to significantly higher skewness for any value of the impurity concentration. In the random case, the skewness is small but non-zero. This reflects an inherent bias in the system that arises from the chirality of the ground state. In other words, there is a small preference for current loops of one chirality over another. However, this is a much weaker effect than that induced by correlations in disorder. 

Measurements with $NV$ centres are particularly suited to finding statistical properties of field distributions. As shown above, correlations in the field distribution reflect underlying correlations of the disorder realization. This makes the case for an additional tuning handle in the form of disorder-correlations. This provides for an additional layer of measurable signatures in a chiral superconductor. 
With adatoms on a candidate material, these measurements can reveal the character of the underlying superconductor. In particular, the contrast between random and correlated disorder serves as a clear signature of chiral 
superconductivity. 

\section{Discussion}
A fascinating consequence of chiral superconductivity is the presence of persistent currents along the sample edge. Despite sustained efforts over decades, such currents have not been observed in experiments. We discuss an alternative in the form of disorder-induced current. Our results show that (a) impurities generically induce local currents in a chiral superconductor and (b) correlated disorder can lead to long ranged currents within the sample. These properties offer significant advantages over an edge current setup. At a simple edge, the bulk chirality serves as a biasing field that drives a current. This bias is in-built and not tuneable. In contrast, a disorder potential provides an `external' biasing field that is both tuneable and flippable. This allows for disorder-currents to be tuneable in orientation, strength and spatial extent. As a general principle, currents driven by an external biasing field are easier to detect. This is exemplified by the recent detection of an edge current in MoTe$_2$\cite{Wang2020}, where an external current was used to bias the edge.

Our analysis is based on a toy model for chiral $p_x \pm i p_y$ superconductivity on a square lattice. However, our results are not model-dependent. Rather, they reflect a generic symmetry-allowed phenomenon -- an impurity in a chiral 
(quasi-)two-dimensional system gives rise to a localized circulating current. This idea can be traced back to studies on integer quantum Hall systems with plateau-shaped potentials\cite{Ford1994}. More recently, impurity-induced currents have been demonstrated in a Chern insulator, a non-superconducting phase that breaks time-reversal symmetry\cite{Jha2017}. Extrapolating this physics to a chiral superconductor requires careful consideration for the following reasons: (a) The superconducting order parameter can re-organize itself to generate `screening' currents. This can be viewed as a local manifestation of the Meissner effect, where the superconductor resists any magnetic flux generated by impurity-currents. (b) The current in a chiral superconductor is carried by `quasiparticles' - a mixture of electrons and holes. If the electron and hole contributions are comparable, the net charge current may vanish despite there being a robust quasiparticle current. In this light, the calculations presented in this article clarify that (i) an impurity in a chiral superconductor generates a localized current loop, and (ii) its net charge current is non-zero, generating a magnetic field.  

Impurity-generated currents have been previously reported in chiral superconductors by Guo et. al.\cite{Guo2017}. However, their approach was not fully self-consistent as the vector potential was ignored. Our study addresses this issue by fully accounting for vector potentials and screening currents.  We have used an improved Bogoliubov-deGennes scheme\cite{Zhu2016}, where the order parameters \textit{and} the vector potential are determined self-consistently. As far as we are aware, this has only been achieved previously in certain simple cases that are amenable to semi-analytic methods. A classic example is the calculation of the profile of a single vortex in a superconductor\cite{Tinkham}. A second, more pertinent, example is edge current in the context of Sr$_2$RuO$_4$\cite{Furusaki2001}. In both these examples, symmetry reduces the vector potential and the order parameter into functions of one variable. Both can be calculated self-consistently in this effective one-dimensional problem. Such approaches are inherently limited and cannot be extended beyond idealized cases. Our algorithm, based on Eq.~\ref{eq.Avec}, is suitable for a wider array of problems. For example, it can be used to determine field distributions in vortices and vortex lattices. 

Our approach complements earlier studies on impurities in chiral superconductors. In particular, Refs.~\onlinecite{Kaladzhyan2016} and \onlinecite{Kreisel2021} present detailed studies of impurity-induced in-gap states. They further discuss periodic arrangements of impurities, where bound states hybridize to form bands that can have topological character. Here, we do not focus on bound states or quasiparticle bands. Rather, we discuss currents induced in the superconducting ground state and in turn, the magnetic fields that they generate. In the discussion above, we have restricted our attention to weak disorder and zero temperature, a situation in which impurity-induced bound states do not play a role. In future studies on realistic systems, their effects can be taken into account. 

Going beyond isolated impurities, we have discussed currents generated by two disorder schemes. We show that correlated disorder can lead to long-ranged current loops within the sample. This is reflected in the magnetic field distribution, providing more complex signatures of chirality.  
Our two disorder schemes are inspired by studies of impurities\cite{Balatsky2006} and adatoms. In several contexts, adatoms have been shown to bring out interesting properties. For instance, adatoms on graphene have been shown to give rise to superconductivity\cite{Profeta2012,Ludbrook2015}.

\section{Acknowledgements}
The authors acknowledge use of the Nandadevi computing cluster 
facility at The Institute of Mathematical Sciences (IMSc), Chennai, India.

\bibliographystyle{apsrev4-1}
\bibliography{chiral_pwave_impurity}
\appendix

\end{document}